\documentclass[12pt]{iopart}
\usepackage{graphicx}
\usepackage{hyperref}

\begin{document}

\title[Functional uncertainty quantification for isobaric MD and defect energies]{Functional uncertainty quantification for isobaric molecular dynamics simulations and defect formation energies}
\author{Samuel Temple Reeve$^1$, Alejandro Strachan$^2$\footnote{Corresponding author}}

\address{$^1$ Materials Science Division, Lawrence Livermore National Laboratory, Livermore, California, 94550, USA}
\address{$^2$ School of Materials Engineering and Birck Nanotechnology Center, Purdue University, West Lafayette, Indiana, 47907 USA}
\ead{strachan@purdue.edu}
\vspace{10pt}
\begin{indented}
\item[]March 2019
\end{indented}

\begin{abstract}
Functional uncertainty quantification (FunUQ) was recently proposed to quantify uncertainties in models and simulations that originate from input functions, as opposed to parameters. This paper extends FunUQ to quantify uncertainties originating from interatomic potentials in isothermal-isobaric molecular dynamics (MD) simulations and to the calculation of defect formation energies. We derive and verify a computationally inexpensive expression to compute functional derivatives in MD based on perturbation theory. We show that this functional derivative of the quantities of interest (average internal energy, volume, and defect energies in our case) with respect to the interatomic potential can be used to predict those quantities for a different interatomic potential, without re-running the simulation. The codes and scripts to perform FunUQ in MD are freely available for download. In addition, to facilitate reproducibility and to enable use of best practices for the approach, we created Jupyter notebooks to perform FunUQ analysis on MD simulations and made them available for online simulation in nanoHUB. The tool uses cloud computing resources and users can view, edit, and run end-to-end workflows from a standard web-browser without the need to need to download or install any software.  
\end{abstract}

\vspace{2pc}

\submitto{\MSMSE}

%


\section{Introduction}

The use of modeling and simulation in materials science and engineering (MSE) is growing at a fast pace. Beyond its traditional application to explain the underlying mechanisms that govern material behavior, modeling and simulations have been more recently recognized as key components in materials discovery, optimization and deployment. Integrated Computational Materials Engineering (ICME) is a well-established subfield within MSE \cite{Allison2006} and the US Materials Genome Initiative (MGI) \cite{USNSTC2011}, together with many similar efforts around the world, have resulted in an expanding set of resources, including cyber-infrastructure for simulation \cite{Klimeck2008}, data \cite{Jain2013}, and models \cite{Tadmor2013},  collectively deemed ``materials data infrastructure" \cite{Warren2018}. In order for these data and models (both physics-based and machine-learned) to be useful in decision making, the uncertainties associated with the predictions need to be assessed and quantified. In the case of experimental data, sample pedigree, test procedure, data analysis methods, and the associated uncertainties \cite{sankararaman2013bayesian} need to be documented and quantified. In the case of models and simulation codes, verification and validation tests must be designed and carried out before they are used for science or engineering applications \cite{oberkampf2010verification}. Verification seeks to answer the question: am I solving the equations correctly? That is, it focuses on the accuracy with which the equations in the model are solved. On the other hand, validation seeks to answer the question: am I solving the right equations? That is, validation seeks to establish to what degree the model represents the physical system and process of interest. The goal of validation is to quantify the discrepancy between the model's predictions and experimental results for a specific quantity of interest (QoI). A necessary step for both verification and validation of models and simulation codes is uncertainty quantification (UQ), which seeks to identify and characterize the sources of uncertainties in simulations and in experiments, see for example \cite{sankararaman2011uncertainty}. 

The sources of uncertainty in materials models have disparate sources, from numerical issues and poorly characterized input parameters or boundary conditions, to approximate constitutive laws, physics, and unknown processes. Numerical sources of uncertainty can be quantified via simulation refinement (e.g. finer grids in continuum simulations or reduction in integration timestep in molecular dynamics (MD)), by comparing the result of a simulation with an analytical simulation, or by checking the accuracy to which the simulation satisfies symmetries of the underlying model (e.g. energy and momentum conservation in adiabatic MD). Often, the dominant source of uncertainty and discrepancy with experiments is the constitutive law used in the model, either its form or its parameters. Uncertainties in constitutive models are best assessed during calibration, using probabilistic approaches like Bayesian methods \cite{kennedy2001bayesian}. While such rigorous approaches are not widely used in materials modeling (most calibration efforts result in a single set of {\it optimal} parameters with unknown uncertainties) propagating uncertainties in constitutive models is clearly of great importance. In the case of uncertain input parameters, a variety of techniques exist for uncertainty propagation, including non-intrusive collocation techniques and intrusive, sometimes more efficient, approaches \cite{xiu2003modeling, adams2009dakota, hunt2015puq}. A related, and particularly challenging, problem in the field of materials is UQ across scales \cite{koslowski2011uncertainty, Chernatynskiy2013}. The focus of this paper is on functional UQ, FunUQ: the quantification of uncertainties associated with constitutive laws themselves and not just their parameters. This enables, for example, quantifying the effect of a change in the functional form of a constitutive equation for a specific QoI. 

FunUQ starts by computing the functional derivative of the QoI with respect to the input function: $\frac{\delta Q[f]}{\delta f}(z)$; this distribution quantifies how much the input function, at each value of its independent variable ($z$), matters for the overall prediction. We will show that with this information one can quantify the uncertainties introduced in the QoI if the uncertainties in the input function are known or can be estimated. It also enables correcting the prediction if a more accurate constitutive law become available, without re-running the simulation. Section 2 provides background for FunUQ and lists possible applications as related to MD. Section 3 extends FunUQ to the isothermal-isobaric ensemble and to the calculation of uncertainties in additional properties. Section 4 discusses best practices and introduces a cloud-based tool in nanoHUB which allows anyone to run FunUQ for MD. Finally, conclusions are drawn in Section 5.

\section{Functional uncertainty quantification} \label{sec:fuq}
\subsection{Functional derivatives and functional sensitivity}

Any predicted QoI originating from a model can be formally written as a function of the input parameters \(\{P_i\}\) and its input functions \(\{f_j\}\): 
\begin{equation}  \label{eq:QoI}
	Q=Q(\{P_i\}, \{f_j(\{z_m\}, \{P_n\})\}) 
\end{equation}
where the input functions depend on parameters $\{P_n\}$ and have $\{z_m\}$ as independent variables. While most UQ efforts seek to quantify how the input parameters $\{P_n\}$ and $\{P_i\}$ affect $Q$, FunUQ focuses on the dependence of $Q$ on the functions $\{f_j\}$. For simplicity, and without loss of generality, let us assume that $Q$ depends on a single function: $Q[f]$. In this case, $Q$ is a functional that maps the space of functions to real numbers. To be more concrete, one example for MD is where $Q$ is the total potential energy, $\{P_i\}$ includes the overall thermodynamic state, $f_j$ is the interatomic potential, $z_m$ is the atomic separation distance, and $\{P_n\}$ are the potential parameters. 

The sensitivity of $Q$ with respect to $f$ can be assessed via its functional derivative, defined in terms of the functional differential as:
\begin{equation} \label{eq:FD0}
  \delta Q[f] = \int {\frac{\delta Q[f]}{\delta f}(z) \delta f(z) dz} 
\end{equation} 
where $\delta f$ is any variation of $f$. The functional derivative is a distribution of the independent variable $z$ and quantifies how $Q$ depends on each the value of $f$ for each possible value of $z$. This is, perhaps, more clearly seen from the differential definition of functional derivative:
\begin{equation} \label{eq:FD1}
\frac{\delta Q[f]}{\delta f(z)}(z_i) = \lim_{\epsilon \to 0} \frac{Q[f(z) + \epsilon\cdot\delta(z-z_i)] - Q[f(z)]}{\epsilon}.
\end{equation}

We use this definition to compute functional derivatives in MD simulations in an explicit, brute force manner to verify a computationally efficient approach based on perturbation theory.

\subsection{Functional uncertainty propagation and error correction}

Armed with the functional derivative, we now discuss its various possible uses within FunUQ. 

{\it Uncertainty propagation}. 
The functional derivative is actually a first order sensitivity and can be used to propagate functional uncertainties in the input function, $\Delta f(z)$, across the model:
\begin{equation} \label{eq:prop}
\Delta Q_{prop} = \int \left| \frac{\delta Q[f]}{\delta f(z)}(z_i) \right| \Delta f(z) dz 
\end{equation} 
This is a generalization of the multi-variate error propagation equation to the space of functions. We note that the uncertainties in constitutive models are often non-constant and depend on the independent variable. For example, the mechanical response of materials is often accurately known for small deformations and interatomic potentials are better determined near equilibrium. Equation \ref{eq:prop} shows that it is the product of the functional sensitivity and functional uncertainty that matters and that the functional derivative provides information on the conditions (in terms of the independent variable) that determines importance for the problem at hand. We believe that this application of FunUQ is particularly interesting in conjunction with Bayesian model calibration (see for example Ref.~\cite{frederiksen2004bayesian}), or approaches that involve determining the Pareto front in multi-objective cost functions \cite{stobener2014pareto}. Such calibration efforts result in a distribution of constitutive laws (potentials), varying functional forms, parameters, or both; with Eq.~\ref{eq:prop} FunUQ can efficiently propagate these uncertainties across the model.

{\it Error correction or reduction}. 
Imagine a simulation carried out with a low-fidelity (maybe computationally efficient) constitutive law, $f(z)$, and the existence of a higher-fidelity function, $g(z)$. The predicted QoI with $f(z)$ can be corrected, to first order, using the functional derivative:
\begin{equation} \label{eq:correct}
\Delta Q_{corr} = \int \frac{\delta Q[f]}{\delta f}(z) \left( g(z)-f(z) \right) dz.
\end{equation}
Section \ref{sec:fuqmd} will exemplify the application of Eq.~\ref{eq:correct} in MD simulations.

{\it Ranking possible evaluations of a high-fidelity model}. 
Consider a situation where a high-fidelity model, $g(z)$, is available to be used as input to the simulation but its computational intensity prohibits its use throughout the simulation \cite{barton2011call}. An example of this situation would be a polycrystalline plasticity model as input to a coarse grain finite element simulation \cite{barton2008embedded}. One can envision using a low-fidelity, computationally inexpensive model, $f(z)$, and use FunUQ to rank where and when to use the high-fidelity model for maximum impact. We see from Equation \ref{eq:correct} that the product of functional derivative and the difference between models $(g(z)-f(z))$ determines how much each evaluation of $g(z)$ contributes to correcting the predicted QoI. Thus, the values of $z$ for which this product is the highest should be considered highest priority. Of course, this quantity depends on $g(z)$, which we are trying to avoid evaluating. Ref.~\cite{Strachan2013} discusses how discrepancy modeling together with FunUQ can used to address this issue.

\section{FunUQ for MD: extension to isobaric conditions and defect formation energies} \label{sec:fuqmd}

MD simulations are playing an increasingly important role in science and engineering. In the field of materials, MD plays a critical role in multiscale modeling by connecting the fundamental interactions between atoms, described from first principles by quantum mechanics, and  phenomena resulting from the collective action of a large number of atoms. MD has contributed to, among other applications, our understanding of the thermal, mechanical, and chemical response of materials under extreme conditions \cite{kadau2002microscopic, shen2016nanosecond, wood2015ultrafast}, mechanical response \cite{zepeda2017probing,Reeve2017lowstiffness}, the nature of glasses \cite{chen2015fractal}, and devices of technological relevance \cite{onofrio2015atomic, liu2016intrinsic}. While uncertainties in MD simulations can originate from a plethora of factors, including simplified microstructures, short simulation time scales, data analysis, lack of ergodicity, etc. \cite{Patrone2015, alzate2018uncertainties, Zhou2017}, MD makes only two fundamental approximations: i) the use of classical mechanics to describe the motion of atoms (neglecting quantum effects) and ii) the use of an approximate description of atomic interactions. While the effect of quantum ionic dynamics is relatively well understood for several applications \cite{berens1983thermodynamics}, the effect of the errors in atomic interactions on the various properties of interest are poorly characterized. This is particularly true when interatomic potentials (force fields) are used in the simulations, but even the use of electronic structure calculations to compute forces (\textit{ab initio} MD) results in uncertainties. Thus, it is not surprising that significant efforts have been devoted to the development of accurate interatomic potentials for various materials classes and, in recent times, the quantification of uncertainties originating from the potential. The vast majority of work to date explores parametric uncertainty within a given potential functional form  \cite{Rizzi2012, rizzi2013uncertainty} or performs systematic comparisons of potentials  \cite{alzate2018uncertainties, Becker2013, Hale2018}. There is also important work investigating Bayesian ensembles of potentials \cite{frederiksen2004bayesian, Longbottom2019}, which is still within a single functional form. Our work on FunUQ for MD simulations seeks to directly establish how the functions used as input affect the prediction of various quantities of interest. 

FunUQ was previously applied to MD simulations in the canonical ensemble \cite{Reeve2017funuq} to assess how average internal energy and pressure depend on the input interatomic potential for liquid and solid samples at various conditions. In this prior work, we used the Lennard-Jones two-body potential as the low-fidelity model and created a family of synthetic modified potentials, considered as high-fidelity. A key challenge addressed in Ref.~\cite{Reeve2017funuq} was the derivation and validation of a computationally efficient approach to compute functional derivatives in MD simulations. We derived an approach based on perturbation theory that adds little cost to the base MD simulations and verified it against explicit calculations of the functional derivative based on Eq.~\ref{eq:FD1}. The latter, brute force calculations involve perturbing the two-body potentials with narrow Gaussian distributions at various interatomic distances and re-running the simulations to obtain the QoI for the modified potentials. This enables evaluation of Eq.~\ref{eq:FD0} numerically. Needless to say, this approach is computationally intensive as it requires running the simulation of interest multiple times, for various perturbations and for several values of the independent variable. In contrast, the perturbative approach, described in detail in Section \ref{sec:perturb}, requires evaluating the perturbed potential along the original MD trajectory. This increases the total computational cost by a negligible amount, on the order of seconds, compared to many minutes for small simulations, to huge numbers of core-hours for large-scale simulations. 

We showed that with FunUQ we could predict both QoIs (internal energy and pressure) for the various samples and conditions for a family of modified interatomic potentials without re-running the original MD simulation (carried out with the Lennard-Jones potential). This was done by comparing the FunUQ result with independent MD simulations using the family of synthetic high-fidelity models. The only cases where the approach resulted in inaccurate corrections were those in which the low- and high-fidelity potentials exhibited significant differences in the configuration space explored (e.g. cavitation in a liquid with only one potential due to significantly higher attraction). In such cases, the perturbative expression fails as the trajectory of the low-fidelity potential was not a good representation of that of the high-fidelity one.

\subsection{System of interest}

In this paper we extend our previous work under isothermal-isochoric (NVT ensemble) conditions to isothermal-isobaric (NPT ensemble) conditions. We generalize the perturbative approach to compute functional derivatives isobaric conditions and verify it against explicit calculations. We also show the ability of FunUQ to provide accurate corrections for not just average internal energy and volume, but also defect formation energies,
from one pairwise interatomic potential to another, without re-running the simulation. 

\begin{figure}
  \makebox[\textwidth][c]{\includegraphics[width=3.3in]{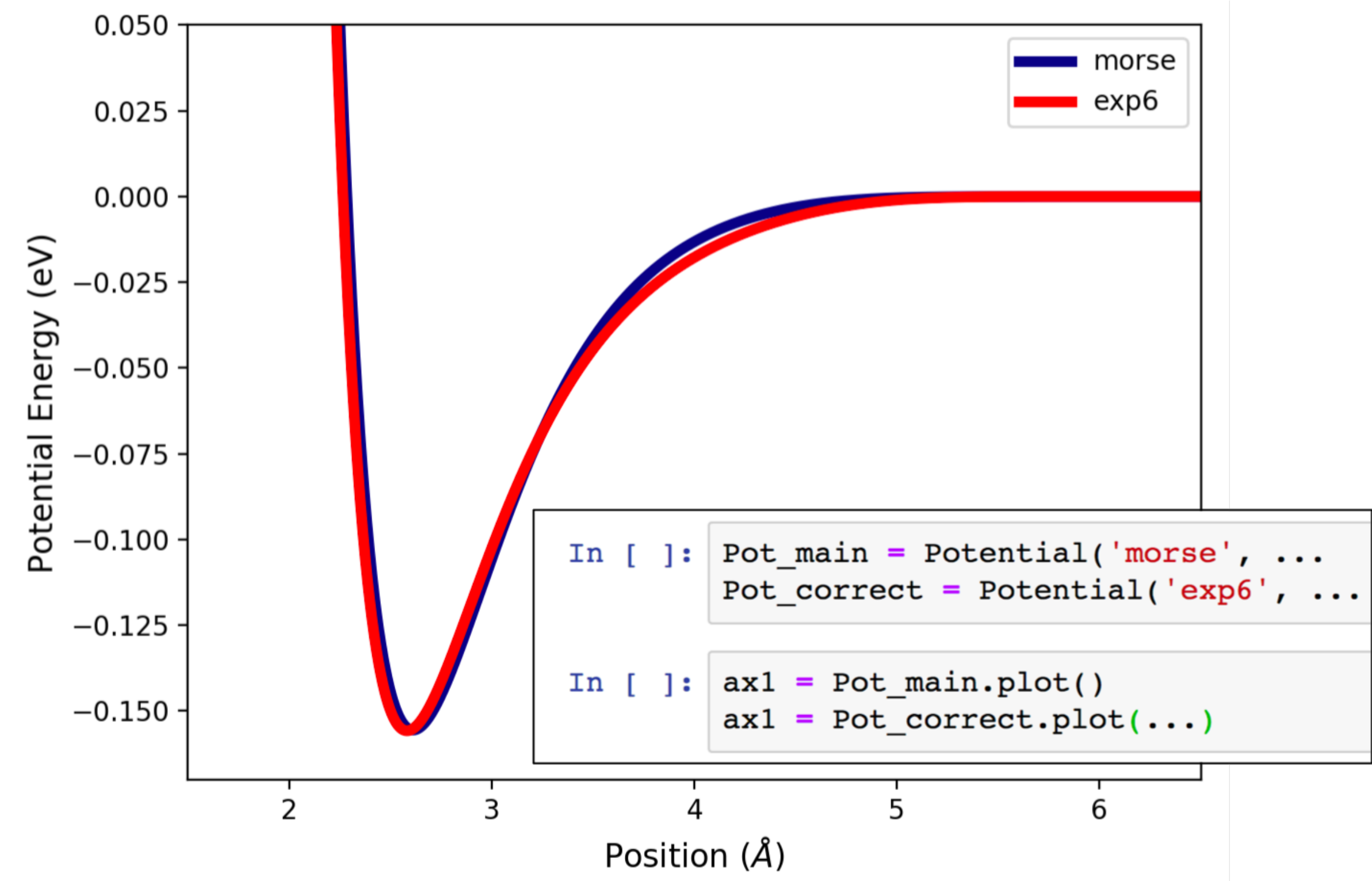}}
  \caption{Input interatomic potentials (Morse and exponential-6), with inset showing the main commands to generate this plot, as described in Section \ref{sec:nH}.}
  \label{fig:pot}
\end{figure}

We use a pairwise Morse potential as the low-fidelity model and an exponential-6 (Exp-6) potential as the high-fidelity model. Both potentials use an exponential function to represent short-range repulsion, but use different forms (exponential and a power law, respectively) to describe the attractive component. These two potentials are compared in Figure \ref{fig:pot}, smoothed to fourth order, tabulated, and visualized with the online tool described in the Section \ref{sec:nH}. We note that it is not necessarily true that the Exp-6 potential is more accurate than Morse; however, we use these two widely-used forms (rather than a synthetic modified potential) as an example of the method to assess different functional forms. All MD simulations in this paper have been carried out using the parallel code LAMMPS, from Sandia National Laboratories \cite{plimpton1995fast}. Standard MD choices included 1 fs timestep and Nos{\'e}-Hoover coupling constants of 0.1 for the thermostat and 1.0 for the barostat. 

\subsection{Numerical and perturbative approach to functional derivatives} \label{sec:perturb}

Prior work on FunUQ for MD described a brute force, compute intensive method to calculate the functional derivative, requiring multiple independent simulations for each position and size of the functional perturbation \cite{Reeve2017funuq}. Out of practical necessity, a computationally efficient, perturbative approach to calculate functional derivatives was developed. Here we extend our approach to isothermal-isobaric simulations and highlight underlying details. 

Central to the calculation of the functional derivative from MD simulations using Equation \ref{eq:FD1} are statistical averages of the QoIs from simulations that use the base interatomic potential with the addition of localized perturbations. The perturbed Hamiltonian $H$ can be written as the sum of the original Hamiltonian with the low-fidelity potential, $H_0$, and a perturbation $H'$. We take this perturbation to be a narrow Gaussian $\phi'(r) = g(r-r_0)$ that is added to the pairwise potential.

Under isothermal-isobaric conditions, the average of a QoI with the perturbed Hamiltonian is \cite{Frenkel2002}:
\begin{equation} \label{eq:ensavg}
\langle Q \rangle_H =
        \frac{\int Q \cdot e^{-\beta \left(H + P V \right)} }
        {\int              e^{-\beta \left(H + P V \right)}}
        = 
        \frac{\int (Q_0+Q') \cdot e^{-\beta \left(H_0 + P \cdot V_0 \right)} \cdot e^{-\beta \left( H' + P \cdot V' \right)}}
        {\int e^{-\beta \left(H_0 + P \cdot V_0 \right)} \cdot e^{-\beta \left( H' + P \cdot V' \right)}},
\end{equation}
where $\beta$ is the inverse thermodynamic temperature, the integrals are over volume and phase space ($d^{3N}r ~ d^{3N}p$), and the factorization of the exponential is enabled by the additive nature of the perturbation.

Equation \ref{eq:ensavg} can be rearranged, by multiplying both denominator and numerator by factors of \(\int e^{-\beta H_0}\), resulting in:
\begin{equation} \label{eq:perturbative}
\langle Q \rangle_H = 
\frac{\langle (Q_0+Q') \cdot e^{-\beta (H'+P \cdot V')} \rangle_{H_0} }{\langle e^{-\beta (H'+P \cdot V')} \rangle_{H_0}}.
\end{equation}

Equation \ref{eq:perturbative} shows that a canonical average over the perturbed Hamiltonian can be written as a ratio between canonical averages over the unmodified potential. The equation also highlights that the contribution to the QoI from the perturbation, $Q'$, must be included to obtain the final, correctly weighted FunUQ correction. 

For example, if the quantity of interest is the internal energy, then:
\begin{equation} \label{eq:PEall}
Q' = \sum_{i<j} \phi'(r_{ij}) =  \sum_{i<j} g(r_{ij}-r_0)
\end{equation}
where $\phi'(r)$ is the perturbation potential, taken to be a localized Gaussian, $g$. This is the sum of the energy contributions from all pairs of atoms, \textit{from the perturbation}, and should be calculated for all the values ($r_0$) of the independent variable used to describe the functional derivative. 
In practice, this calculation can be done during the base MD simulation with the unmodified potential or as a post-processing step, using only the perturbations. This later calculation can be simplified by using the instantaneous radial distribution function (RDF) coordination number obtained from the MD simulation:
\begin{equation} \label{eq:PEcoord}
Q' = \frac{N}{2} \sum_{k} \phi'(r_k) \cdot c(r_k). 
\end{equation}

Thus, a single simulation using the unmodified, low-fidelity potential is required to evaluate the complete functional derivative. Since an isothermal-isobaric MD simulation with the base potential results in configurations with the correct statistics, the averages in \ref{eq:perturbative} can be approximated by averages over the unperturbed MD simulation:
\begin{equation} \label{eq:mdpert}
\langle Q \rangle_H \sim
\frac{\langle (Q_0+Q') \cdot e^{-\beta (H'+P \cdot V')} \rangle_{MD/H_0} }{\langle e^{-\beta (H'+P \cdot V')} \rangle_{MD/H_0}},
\end{equation}
where $\langle \rangle_{MD/H_0}$ denotes a time average over MD simulation(s) with Hamiltonian $H_0$.

 This expression is an extension to isobaric conditions of the expressions derived in Ref.~\cite{Reeve2017funuq} under the canonical ensemble and is similar to free energy methods including thermodynamic integration and free energy perturbation \cite{Chipot2007}. 
 While Eq.~\ref{eq:perturbative} is exact, replacing the ensemble average for a time average in \ref{eq:mdpert} is an approximation and its accuracy will depend both on simulation time and the size of the perturbation. If the two interatomic potentials exhibit poor overlap in the regions of phase space explored, e.g. one potential predicts a crystal and the second one a melt, Eq.~\ref{eq:mdpert} is a poor approximation and would result in erroneous predictions. 
 The use of the RDF to compute the energy of the perturbation is a further numerical approximation and can be systematically improved with finer discretization. A number of practical considerations are necessary to compute accurate functional derivatives using the preceding expressions; these are discussed in Section \ref{sec:bestpractices}, where we also discuss the computational intensity of the calculation.

\subsection{Correcting predicted volume and internal energy} \label{sec:correct1}

To test the ability of FunUQ to correct predictions in the NPT ensemble, we use a monatomic liquid at 1500 K and 1 GPa pressure as the example and average internal energy and volume as the QoIs. All results are presented as computed and visualized within the online tool described in Section \ref{sec:nH}; readers can run the simulations as they read the paper. 

\begin{figure}
  \makebox[\textwidth][c]{\includegraphics[width=3.3in]{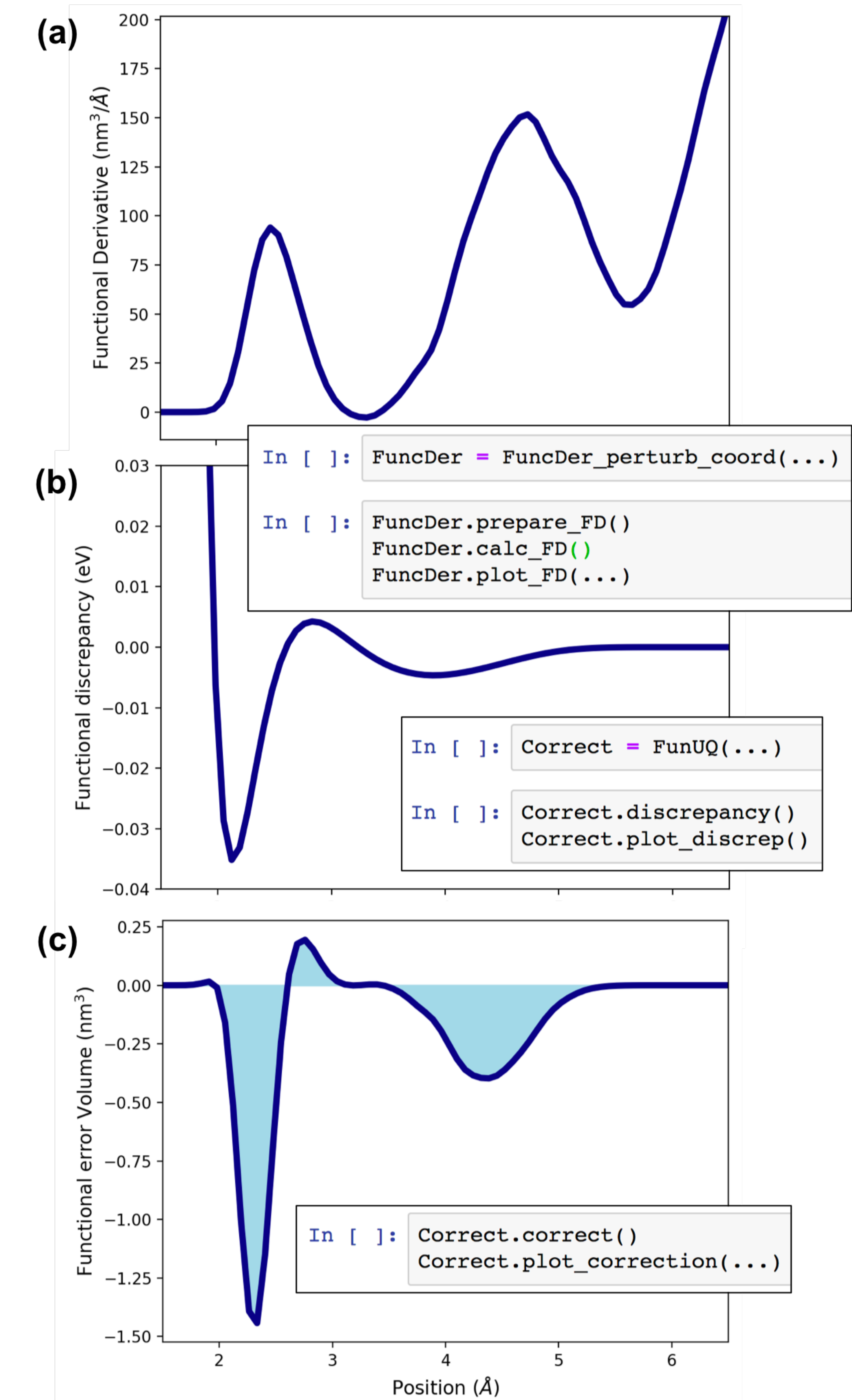}}
  \caption{FunUQ example for volume including (a) functional derivative, (b) functional discrepancy, and (c) functional error, where the integrated error is the total correction. Inset in each are the main commands to generate the plot, as described in Section \ref{sec:nH}.}
  \label{fig:results}
\end{figure}

The main simulation with the low-fidelity Morse potential was run for a total of 2 ns (with 2 replicas), within the NPT ensemble. The functional derivative was computed using perturbations with heights $\pm 1 \cdot 10 ^{-8}$ and $2 \cdot 10 ^{-8}$ and width of 0.1 $\AA$ using the RDF (discretized over 2,000 points) saved during the trajectory every 1 ps at 200 points evenly spaced from zero to the potential cutoff. The discrepancy was calculated as the difference between the two potentials in tabular form (Figure \ref{fig:pot}) and the product between functional derivative and discrepancy integrated to calculate the total correction (Eq.~\ref{eq:correct}). Figure \ref{fig:results} shows these three main steps, as well as snippets of code from Section \ref{sec:nH} for volume as the QoI. 

\begin{figure}
	\makebox[\textwidth][c]{\includegraphics[width=6.3in]{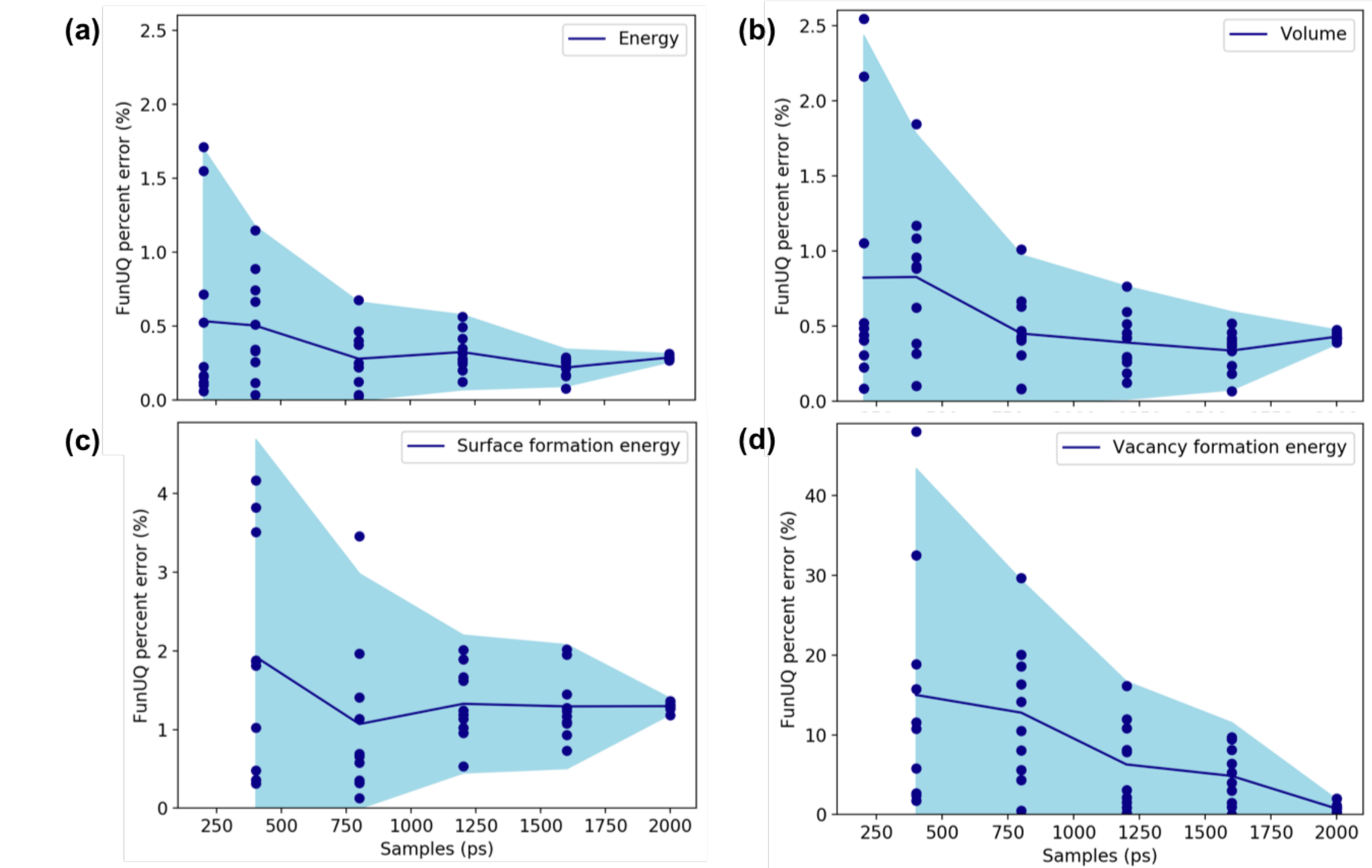}}
	\caption{FunUQ prediction error as a function of MD simulation time used to calculate the functional derivative for (a) internal energy, (b) volume, (c)  surface formation energy, and (d) vacancy formation energy. Points show individual calculations, lines the average for a given sample size, and the shaded region $\pm$ 2 standard deviations.}
	\label{fig:converge}
\end{figure}

In order to assess the accuracy of the FunUQ corrections, we performed corresponding simulations with the Exp-6 potential. The convergence of the FunUQ prediction error is shown in Fig. \ref{fig:converge}; for each, random samples of increasing size were taken from each replica to calculate the functional derivative, repeated 10 times. For simplicity, here we report the average results using the full 2 ns. The difference between the Exp-6 and Morse predictions for average volume is -0.720 nm$^3$ and the FunUQ correction -0.774 nm$^3$, while the original Morse result is 13.9 nm$^3$. Thus, the low fidelity simulation plus correction results in an average error of only 0.43\%. Similarly, the Morse potential predicts an average internal energy of -0.666 ev/atom, the direct simulation difference between Exp-6 and Morse is 0.0627 eV/atom, and the FunUQ correction is 0.0645 eV/atom, resulting in a small average error of 0.29\%. These results are comparable to the correction errors of pressure and energy for NVT systems correcting between the same potentials, 6.5\% and 0.35\%, respectively (also 1500 K, but at 1 atm). Correcting pressure highlights that different cases show different convergence and overall accuracy, particularly when the different input potentials begin to diverge in phase space, here due to significant differences in pressure when constrained to constant volume. Additional simulation time will continue to both slightly reduce the FunUQ error and slightly increase the uncertainty bounds of the prediction.

\subsection{Correcting defect formation energies} \label{sec:correct2}

So far we have demonstrated that FunUQ can be used to quantify how the value of bulk, equilibrium properties change in response to change in the interatomic potential used in MD simulations. In materials science, one is often interested in defect formation energies and we now demonstrate FunUQ for free surface formation energy and vacancy formation energy.

Defect formation energy calculations involve the energy difference between a perfect system and one with the defect. Thus, two functional derivatives are calculated: one for the perfect bulk and one for the defective system. We chose these systems to be solid FCC crystals equilibrated at 300K and 1 atm in the NPT ensemble. For the surface calculation, one direction was made non-periodic resulting in two free surfaces and for the vacancy calculation one atom was removed from the crystal. The inputs for calculation of the functional derivatives matched Section \ref{sec:correct1}. With the energy correction predicted for both systems through FunUQ, the total FunUQ surface formation energy and vacancy formation energy are then calculated through the standard formulas: 
\begin{equation}
  E_{surf-form} = \frac{(E_{surf} - E_{bulk})}{2 A},
\end{equation}
where $A$ is the area of a each of the free surfaces, and 
\begin{equation}
  E_{vac-form} = E_{vac}\frac{(N-1)}{N} - E_{bulk}
\end{equation}
where N is the number of bulk atoms. 

As above, we take Morse simulations as the base and use FunUQ to predict the defect formation energies expected for the Exp-6 potential. Using a total of 2 ns of simulation time (2 replicas) to compute functional derivatives, 
FunUQ predicts a surface formation energy of 892 mJ/m$^2$ and a vacancy formation energy of 2.26 eV. Explicit MD simulations with the Exp-6 potential result in 909 mJ/m$^2$ and 2.23 eV, resulting in average FunUQ errors of 1.3\% and 0.73\% for surface and vacancy formation energies, respectively, showing that FunUQ can be used to correct uncertainties in defect formation energies. 
The convergence of the corrected result with respect to the total simulation time used to compute the functional derivative is shown in Fig. \ref{fig:converge} for the various QoIs.  We note relative errors are higher for the defect formation energy calculations and longer simulation times are needed to obtain acceptable errors.
This is expected, since converging defect formation energies, that involve the difference between two numbers of similar magnitude, is more challenging than bulk properties. Both defect energies converge much more slowly than the bulk values, where the vacancy formation energy, the smallest difference to resolve, shows much more significant errors with small samples.

\section{Reproducibility and numerical aspects of FunUQ} \label{sec:bestpractices}

In this Section we describe numerical aspects of the FunUQ calculations and describe an implementation of FunUQ that can be used together with the MD code LAMMPS.
All the underlying codes and scripts required to perform FunUQ in MD simulations are publicly available via a git repository (\url{https://github.rcac.purdue.edu/StrachanGroup/FunUQ}). In order to simplify the reproduction of published results and to allow other researchers to fully understand FunUQ calculations and judiciously choose FunUQ inputs, we created and deployed a tool for online simulations in nanoHUB. With the ``FunUQ for MD'' tool, available at \url{https://nanohub.org/tools/funuq} \cite{funuqTool}, anyone with an internet connection can perform simulations using cloud computing without downloading or installing any software. 

\subsection{Cloud computing tool for FunUQ in MD} \label{sec:nH}

Built on top of Jupyter notebooks available within nanoHUB, the Python FunUQ module runs LAMMPS simulations (on nanoHUB or remote computing resources), computes functional derivatives and corrections, and plots results. Users need only a nanoHUB account (free), minimal familiarity with Python and Jupyter, and a web browser. 


\begin{figure}
  \makebox[\textwidth][c]{\includegraphics[width=3.3in]{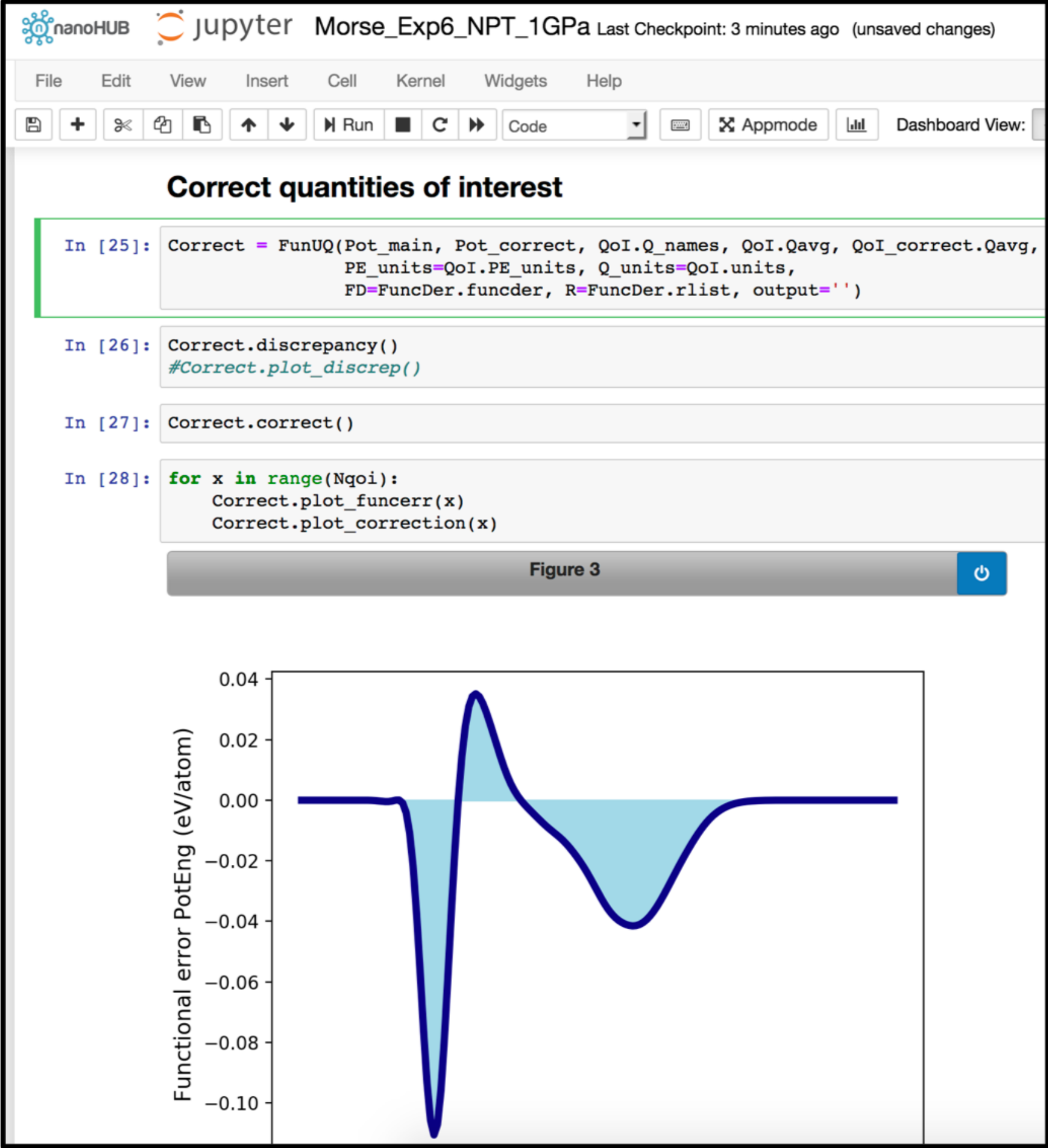}}
  \caption{nanoHUB FunUQ for MD tool snapshot, showing the final steps for error correction.}
  \label{fig:snapshot}
\end{figure}

The tool opens with a simple project manager, keeping track of separate notebooks and simulation folders.
One example notebook for the previous MD publication \cite{Reeve2017funuq} is available, matching results for functional derivative, discrepancy, and error in Fig.~4 and total corrections in Fig.~5 therein. Multiple notebooks are available from this work for all results in Section \ref{sec:fuqmd}. Figure \ref{fig:snapshot} shows one view of the tool, as seen in the user's browser. The code consists primarily of separate Python classes for each of the main parts of the calculation: \texttt{Potential}, \texttt{QuantitiesOfInterest}, \texttt{FuncDer}, and \texttt{FunUQ}. The functional derivatives are a set of inherited classes for the brute force, all-atom perturbative, and RDF based perturbative calculations, where the last is primarily used. Only main details are exposed to the user to run MD, calculate properties, and make plots. 

For users running new cases, a pre-installed notebook is copied to a new project, along with a folder where necessary LAMMPS input and structure files can be uploaded, matching the style of the examples. If needed, the underlying code can be modified within the user's local version.
The examples shown here use interatomic potentials built through the FunUQ code (using LAMMPS \texttt{pair\_style table}); simulations with any feasible LAMMPS \texttt{pair\_style} are also possible if the pair function is written into a table (using LAMMPS \texttt{pair\_write}) for later corrections.

\subsection{Functional derivative calculation}

The calculation of the functional derivative requires specifying the width and height of the Gaussian perturbations, separation between points where the perturbation is centered, and the discretization of the RDF, if used. 

\begin{figure}
	\makebox[\textwidth][c]{\includegraphics[width=6.3in]{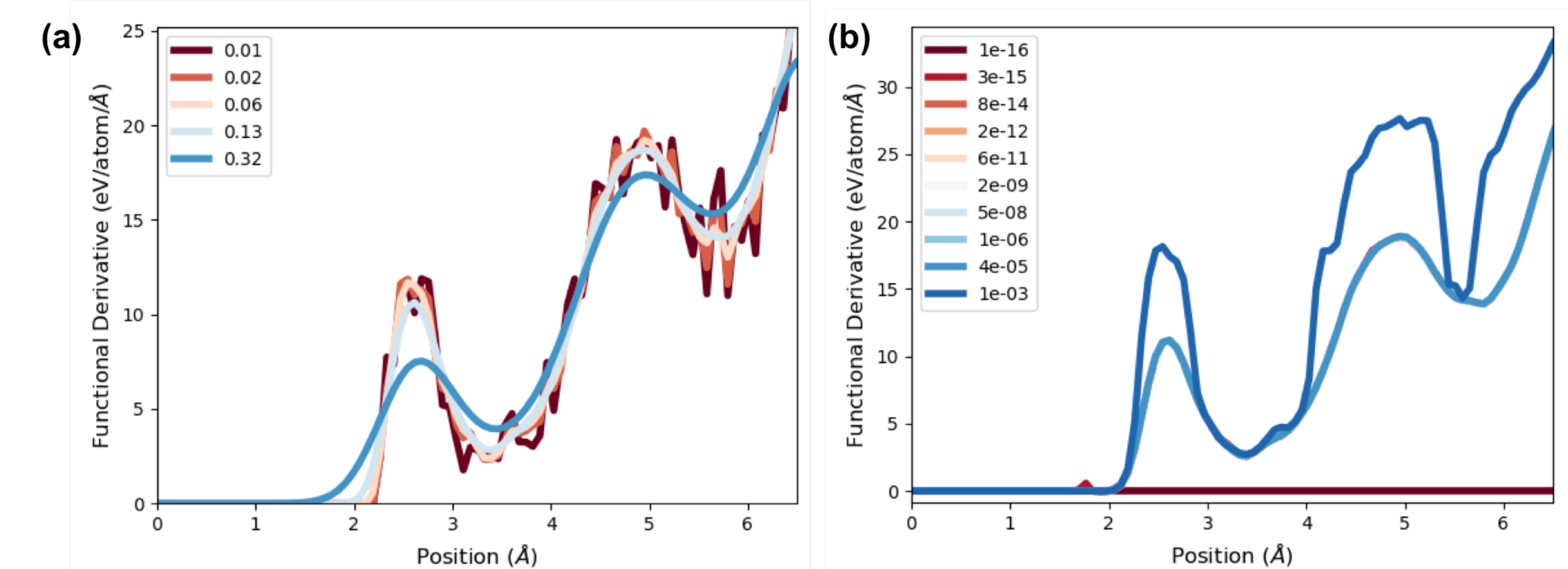}}
	\caption{Exploration of optimal perturbation sizes in functional derivatives for MD (NPT) varying (a) widths and (b) heights.}
	\label{fig:perturb}
\end{figure}

Perturbation widths should be chosen to localize the perturbation, minimizing the spreading or smoothing out of the functional derivative while ensuring a smooth, measurable response, Fig.~\ref{fig:perturb}(a). A relatively small window of perturbation widths is useful, where less than one order of magnitude smaller than that used (0.1) has significant noise, while larger quickly loses local accuracy of the sensitivity. Choice of perturbation heights balances efficiently calculating the exponential weights and ensuring they are large enough to produce a noticeable effect, Fig.~\ref{fig:perturb}(b). We find a very broad range where the resulting functional derivative is insensitive to the choice of perturbation heights; eventually a large enough perturbation becomes difficult to converge, while extremely small perturbations cannot be resolved numerically. Linearity with respect to all perturbations of a given value of the independent variable should also be confirmed. 
The perturbation sizes used in this work are significantly smaller ($10^{-8}$) than in the previous work ($10^{-4}$)\cite{Reeve2017funuq}, helping to reduce the total amount of simulation time to convergence.
Variation of these parameters is demonstrated using the NPT system from Section \ref{sec:correct1}.

\begin{figure}
	\makebox[\textwidth][c]{\includegraphics[width=3.3in]{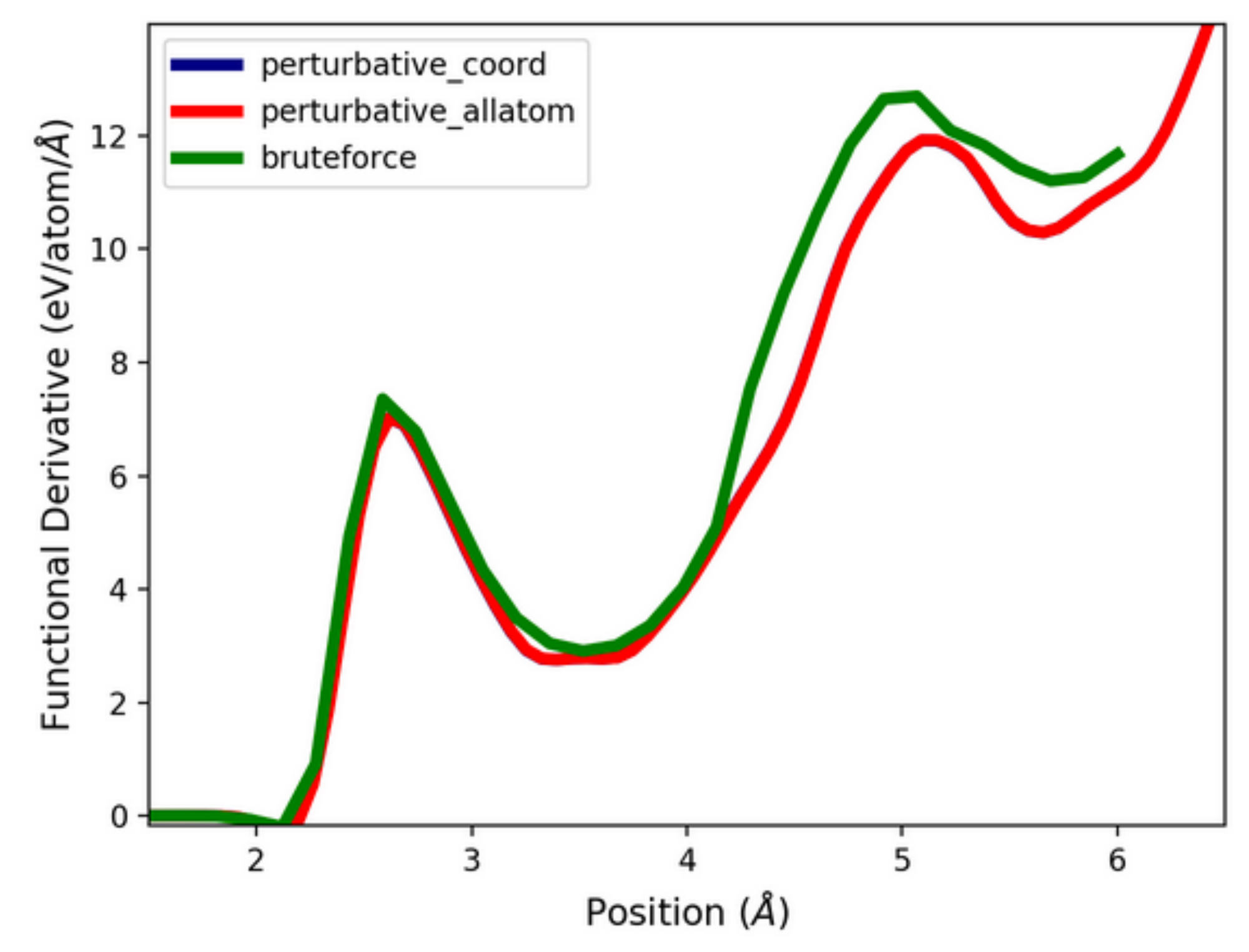}}
	\caption{Comparison of functional derivatives for MD (NVT) using the perturbative method with instantaneous RDF (coordination) and all atoms, as well as the brute force method.}
	\label{fig:FDmethod}
\end{figure}

The use of RDF to calculate the contribution from the perturbations in Equation \ref{eq:perturbative} modestly reduces both storage and computation of the functional derivative compared to evaluating the expression over the ensemble of atoms. The results for well converged functional derivatives with the instantaneous RDF or the all-atom method are indistinguishable in Fig.~\ref{fig:FDmethod}. The brute force functional derivative generally matches, but takes significant amounts of simulation time to converge.
Comparison of these methods is demonstrated using the NVT system from Section \ref{sec:correct1}.

Convergence with a sufficiently long MD simulation is imperative for a successful calculation, checked for each case here and notably different for each property and system. Convergence for this work was done using the total correction, compared with the reference direct simulation results. The total integrated functional derivative could also be used to check convergence since it is the exponential weighting that needs to converge.
Finally, the number of points in the functional derivative should be large enough to resolve all features and ensure smoothness, particularly for crystalline systems with more abrupt changes.

 The computational intensity of the functional derivative calculation can be assessed in the following way. In order to sample the independent variable, we use $N_{r_0}=200$ values of $r_0$ as the center of the perturbation and for each $r_0$ we compute $N_h=4$ perturbation heights. Each of these evaluations is comparable in computational intensity to the evaluation of the potential energy of the base MD simulation.  As with any canonical average, Eq.~\ref{eq:mdpert} does not need to computed at every MD step, since adding highly correlated configurations does not increase the accuracy of the average. If we compute the perturbation every $N_{skip}$ MD steps (1000 in our case), the computational cost of the functional derivative is related to the base MD run by a factor of $N_{r_0} \cdot N_h / N_{skip}$. Thus, this factor is 0.8 in the calculations shown here and could be further optimized. The cost can be significantly reduced by the use of the RDF, especially for large systems. The number of evaluations of the perturbed potential is the number of discretization steps used in the RDF (2,000 in our case) while the number of evaluations of the potential in the base simulation is proportional to the number of atoms times the average number of neighbors within the cutoff radius of the potential.

\subsection{Discrepancy}
Finally, the discrepancy between the two potentials of interest is an important feature in regards to the quality of FunUQ corrections. Within general free energy methods in which one goes from one state to the next in some perturbative manner, the comparison of differences in potential energy distributions of those initial and final states is common to provide understanding of the phase space overlap \cite{Pohorille2010}. This requires using energy distributions from the  trajectories of both potentials, as well as each potential with the opposite trajectory. Without any overlap at all there is no hope in making a good correction; with increasing overlap comes ease of convergence. The details for this calculation were discussed in detail and shown for many cases within Ref.~\cite{Reeve2017funuq}.
Because this requires calculations with the correction potential, which we try to avoid, it is useful to examine trends of total (integrated) discrepancy and the corresponding total overlap to establish bounds of easily converged calculations for many potentials.

Multiple notebooks are included within the nanoHUB tool to explore optimization of functional derivatives and discrepancies.

\section{Discussion and conclusions} \label{sec:conclusion}

In summary, FunUQ enables the quantification of uncertainties originating from input functions used in models and simulations. Central to the method is the functional derivative of the QoI with respect to the input functions, which can be used to propagate uncertainties if uncertainties in the input functions are known or if an ensemble of models are available. In this paper, we extended FunUQ for MD simulations to the isothermal-isobaric ensemble and showed that the method can efficiently predict properties of the exponential-6 potential from a simulation performed with the Morse potential, including both bulk properties and defect energies. 

Other approaches for uncertainty quantification in MD have been developed, including ``reliable" MD which uses generalized intervals to represent the uncertainty in atom positions and velocities with ranges, rather than exact values \cite{Tran2017}. The same method was extended to include an interval for uncertainty in the interatomic potential as well \cite{Tran2016}, in which that uncertainty range was assumed and propagated. While a valuable pursuit with similar goals to FunUQ, the simulations are more computationally expensive and are intrusive to the MD code. 

There are several possible future directions with FunUQ for MD worth exploring. Work to date has focused on equilibrium properties from simulations using pair potentials and in single component systems. With only added bookkeeping, multi-component systems could be handled. Application of FunUQ to more complex and interesting interatomic potentials would be important. We note that the use of the computationally efficient method requires a potential whose perturbation results in an additive contribution to the Hamiltonian. While this restricts the use of this specific approach, several potentials satisfy this condition. For example, most terms used in the simulation of molecular materials are additive \cite{Mayo1990}, as are the two-body part and embedding energy of the embedded atom method \cite{Daw1984} and van der Waals description in reactive potentials like ReaxFF \cite{vanDuin2001}. Even the recent, complex neural network potentials can be partially additively decomposed \cite{Behler2007}. For all of these, the uncertainty due to one portion of the potential could be individually investigated using FunUQ. Finally, additional properties are amenable to FunUQ analysis, from equilibrium fluctuation-based properties to full non-equilibrium simulations. 
 
FunUQ could even be implemented within the now common potential databases, OpenKIM or the NIST Interatomic Potential Repository \cite{Tadmor2013, Becker2013}. This is important to ensure that the error and uncertainties in predictions with FunUQ are acknowledged and understood. Indeed, inclusion of FunUQ within these resources could empower users and developers to make even more useful comparisons between similar potentials. 

Importantly, the results from this paper suggest another new avenue for FunUQ in MD: potential development. Calibration and evaluation of new interatomic potentials requires both many simulations with slightly varying inputs and is concerned with predicted energy, density (volume), defect energies, and similar properties. FunUQ could therefore extend and improve potential development by enabling multiple functional forms to be simultaneously fit and directly compared. Further, the use of Bayesian calibration \cite{frederiksen2004bayesian, Longbottom2019} or Pareto front multi-objective optimization \cite{stobener2014pareto} which result in ensembles of potentials could significantly benefit from FunUQ. A distribution of potential inputs are calibrated for a given system and, to make a prediction of a property, a distribution of those potentials must be run. With FunUQ this becomes computationally tractable, where only one direct simulation is necessary, followed by (or in parallel with) calculation of the functional derivative and propagation through the distribution of parameterized functions. MD simulations with a well defined, propagated uncertainty with a potential ensemble would have a wide impact on the ability to use and trust the resulting predictions. 

We believe FunUQ could be useful in a wide range of  materials simulations where constitutive laws are not known with high accuracy, as well as simulations outside the field of materials. The free energy density functions in phase field methods \cite{Chen2002}, constitutive plasticity laws in mesoscale and continuum simulations \cite{Chaboche2008}, and exchange-correlation functionals in density functional theory \cite{Sousa2007} may be good candidates for this analysis due to many functional forms, ubiquity of the methods, and lack of a ``true" function. We believe the key challenge in the application of FunUQ to these fields will be the efficient evaluation of the functional derivatives.

\ack

This work was partially supported by the U.S. National Science Foundation under contract CBET 1404823.
Part of this work was performed under the auspices of the U.S. Department of Energy by Lawrence Livermore National Laboratory under Contract DE-AC52-07NA27344.
Computational resources of nanoHUB and Purdue are gratefully acknowledged.

\section*{References}
\bibliography{1.bib}

\providecommand{\newblock}{}
\begin{thebibliography}{10}
\expandafter\ifx\csname url\endcsname\relax
  \def\url#1{{\tt #1}}\fi
\expandafter\ifx\csname urlprefix\endcsname\relax\def\urlprefix{URL }\fi
\providecommand{\eprint}[2][]{\url{#2}}

\bibitem{Allison2006}
Allison J, Backman D and Christodoulou L 2006 {\em JOM Journal of the Minerals
  Metals and Materials Society\/} {\bf 58} 25--27

\bibitem{USNSTC2011}
{US National Science and Technology Council} 2011 Materials {{Genome
  Initiative}} for {{Global Competitiveness}}

\bibitem{Klimeck2008}
Klimeck G, McLennan M, Brophy S~P, Adams G~B and Lundstrom M~S 2008 {\em
  Computing in Science \& Engineering\/} {\bf 10} 17--23

\bibitem{Jain2013}
Jain A, Ong S~P, Hautier G, Chen W, Richards W~D, Dacek S, Cholia S, Gunter D,
  Skinner D, Ceder G and Persson K~A 2013 {\em APL Materials\/} {\bf 1} 0--11

\bibitem{Tadmor2013}
Tadmor E~B, Elliott R~S, Phillpot S~R and Sinnott S~B 2013 {\em Current Opinion
  in Solid State and Materials Science\/} {\bf 17} 298--304

\bibitem{Warren2018}
Warren J~A and Ward C~H 2018 {\em JOM Journal of the Minerals Metals and
  Materials Society\/} {\bf 70} 1652--1658

\bibitem{sankararaman2013bayesian}
Sankararaman S and Mahadevan S 2013 {\em Structural Control and Health
  Monitoring\/} {\bf 20} 88--106

\bibitem{oberkampf2010verification}
Oberkampf W~L and Roy C~J 2010 {\em Verification and validation in scientific
  computing\/} (Cambridge University Press)

\bibitem{sankararaman2011uncertainty}
Sankararaman S, Ling Y and Mahadevan S 2011 {\em Engineering Fracture
  Mechanics\/} {\bf 78} 1487--1504

\bibitem{kennedy2001bayesian}
Kennedy M~C and O'Hagan A 2001 {\em Journal of the Royal Statistical Society:
  Series B (Statistical Methodology)\/} {\bf 63} 425--464

\bibitem{xiu2003modeling}
Xiu D and Karniadakis G~E 2003 {\em Journal of computational physics\/} {\bf
  187} 137--167

\bibitem{adams2009dakota}
Adams B~M, Bohnhoff W, Dalbey K, Eddy J, Eldred M, Gay D, Haskell K, Hough P~D
  and Swiler L~P 2009 {\em Sandia National Laboratories, Tech. Rep.
  SAND2010-2183\/}

\bibitem{hunt2015puq}
Hunt M, Haley B, McLennan M, Koslowski M, Murthy J and Strachan A 2015 {\em
  Computer Physics Communications\/} {\bf 194} 97--107

\bibitem{koslowski2011uncertainty}
Koslowski M and Strachan A 2011 {\em Reliability Engineering \& System
  Safety\/} {\bf 96} 1161--1170

\bibitem{Chernatynskiy2013}
Chernatynskiy A, Phillpot S~R and LeSar R 2013 {\em Annual Review of Materials
  Research\/} {\bf 43} 157--182

\bibitem{frederiksen2004bayesian}
Frederiksen S~L, Jacobsen K~W, Brown K~S and Sethna J~P 2004 {\em Physical
  Review Letters\/} {\bf 93} 165501

\bibitem{stobener2014pareto}
St\"obener K, Klein P, Reiser S, Horsch M, K\"ufer K~H and Hasse H 2014 {\em
  Fluid Phase Equilibria\/} {\bf 373} 100--108

\bibitem{barton2011call}
Barton N~R, Bernier J~V, Knap J, Sunwoo A~J, Cerreta E~K and Turner T~J 2011
  {\em International Journal for Numerical Methods in Engineering\/} {\bf 86}
  744--764

\bibitem{barton2008embedded}
Barton N~R, Knap J, Arsenlis A, Becker R, Hornung R~D and Jefferson D~R 2008
  {\em International Journal of Plasticity\/} {\bf 24} 242--266

\bibitem{Strachan2013}
Strachan A, Mahadevan S, Hombal V and Sun L 2013 {\em Modelling and Simulation
  in Materials Science and Engineering\/} {\bf 21} 065009

\bibitem{kadau2002microscopic}
Kadau K, Germann T~C, Lomdahl P~S and Holian B~L 2002 {\em Science\/} {\bf 296}
  1681--1684

\bibitem{shen2016nanosecond}
Shen Y, Jester S~B, Qi T and Reed E~J 2016 {\em Nature Materials\/} {\bf 15} 60

\bibitem{wood2015ultrafast}
Wood M~A, Cherukara M~J, Kober E~M and Strachan A 2015 {\em The Journal of
  Physical Chemistry C\/} {\bf 119} 22008--22015

\bibitem{zepeda2017probing}
Zepeda-Ruiz L~A, Stukowski A, Oppelstrup T and Bulatov V~V 2017 {\em Nature\/}
  {\bf 550} 492

\bibitem{Reeve2017lowstiffness}
Reeve S~T, {Belessiotis-Richards} A and Strachan A 2017 {\em Nature
  Communications\/} {\bf 8} 1137

\bibitem{chen2015fractal}
Chen D~Z, Shi C~Y, An Q, Zeng Q, Mao W~L, Goddard W~A and Greer J~R 2015 {\em
  Science\/} {\bf 349} 1306--1310

\bibitem{onofrio2015atomic}
Onofrio N, Guzman D and Strachan A 2015 {\em Nature Materials\/} {\bf 14} 440

\bibitem{liu2016intrinsic}
Liu S, Grinberg I and Rappe A~M 2016 {\em Nature\/} {\bf 534} 360

\bibitem{Patrone2015}
Patrone P~N, Dienstfrey A, Andrea R, Tucker S and Christensen S 2015 {\em
  Polymer\/} {\bf 87} 1--36

\bibitem{alzate2018uncertainties}
Alzate-Vargas L, Fortunato M, Haley B~P, Li C, Colina C and Strachan A 2018
  {\em Modelling and Simulation in Materials Science and Engineering\/} {\bf
  26} 065007

\bibitem{Zhou2017}
Zhou X and Foiles S~M 2017 Uncertainty {Quantification} and {Reduction} of
  {Molecular} {Dynamics} {Models} {\em Uncertainty Quantification and Model
  Calibration\/} ed Hessling J~P (IntechOpen)

\bibitem{berens1983thermodynamics}
Berens P~H, Mackay D~H, White G~M and Wilson K~R 1983 {\em The Journal of
  Chemical Physics\/} {\bf 79} 2375--2389

\bibitem{Rizzi2012}
Rizzi F, Najm H~N, Debusschere B~J, Sargsyan K, Salloum M, Adalsteinsson H and
  Knio O~M 2012 {\em Multiscale Model. Simul.\/} {\bf 10} 1428--1459

\bibitem{rizzi2013uncertainty}
Rizzi F, Jones R, Debusschere B and Knio O 2013 {\em The Journal of Chemical
  Physics\/} {\bf 138} 194105

\bibitem{Becker2013}
Becker C~A, Tavazza F, Trautt Z~T and Buarque~de Macedo R~A 2013 {\em Current
  Opinion in Solid State and Materials Science\/} {\bf 17} 277--283

\bibitem{Hale2018}
Hale L~M, Trautt Z~T and Becker C~A 2018 {\em Modelling and Simulation in
  Materials Science and Engineering\/} {\bf 26} 055003

\bibitem{Longbottom2019}
Longbottom S and Brommer P 2019 {\em Modelling and Simulation in Materials
  Science and Engineering\/} {\bf 27} 044001

\bibitem{Reeve2017funuq}
Reeve S~T and Strachan A 2017 {\em Journal of Computational Physics\/} {\bf
  334} 207--220

\bibitem{plimpton1995fast}
Plimpton S 1995 {\em Journal of Computational Physics\/} {\bf 117} 1--19

\bibitem{Frenkel2002}
Frenkel D and Smit B 2002 {\em Understanding Molecular Simulation: {{From}}
  Algorithms to Applications\/} ({Academic Press}) ISBN 0-12-267351-4

\bibitem{Chipot2007}
Chipot C and Pohorille A 2007 {\em Free {{Energy Calculations}}\/} (Berlin:
  {Springer}) ISBN 978-3-540-38447-2

\bibitem{funuqTool}
Reeve S~T and Strachan A 2019 Functional {{Uncertainty Quantification}} for
  {{Molecular Dynamics}} \urlprefix\url{https://nanohub.org/resources/funuq}

\bibitem{Pohorille2010}
Pohorille A, Jarzynski C and Chipot C 2010 {\em J. Phys. Chem. B\/} {\bf 114}
  10235--10253

\bibitem{Tran2017}
Tran A~V and Wang Y 2017 {\em Computational Materials Science\/} {\bf 127}
  141--160

\bibitem{Tran2016}
Tran A and Wang Y 2016 {\em ASME. International Design Engineering Technical
  Conferences and Computers and Information in Engineering Conference, Volume
  1A: 36th Computers and Information in Engineering Conference\/}  V01AT02A022

\bibitem{Mayo1990}
Mayo S~L, Olafson B~D and Goddard W~A 1990 {\em J. Phys. Chem.\/} {\bf 94}
  8897--8909

\bibitem{Daw1984}
Daw M~S and Baskes M~I 1984 {\em Phys. Rev. B\/} {\bf 29} 6443--6453

\bibitem{vanDuin2001}
{van Duin} A~C~T, Dasgupta S, Lorant F and Goddard~III W~A 2001 {\em J. Phys.
  Chem. A\/} {\bf 105} 9396--9409

\bibitem{Behler2007}
Behler J and Parrinello M 2007 {\em Physical Review Letters\/} {\bf 98} 146401

\bibitem{Chen2002}
Chen L~Q 2002 {\em Annual Review of Materials Research\/} {\bf 32} 113--140

\bibitem{Chaboche2008}
Chaboche J~L 2008 {\em International Journal of Plasticity\/} {\bf 24}
  1642--1693

\bibitem{Sousa2007}
Sousa S~F, Fernandes P~A and Ramos M~J~a 2007 {\em The Journal of Physical
  Chemistry A\/} {\bf 111} 10439--10452

\end{thebibliography}

\end{document}